\documentstyle[twocolumn,epsfig,aps]{revtex}  

\begin{document} 

\setcounter{footnote}{0}

\title{First Observation of Self-Amplified Spontaneous Emission 
in a Free-Electron Laser at 109 nm Wavelength
\thanks{ 
Dedicated to Bj{\o}rn H.~Wiik, 17.2.1937 -- 26.2.1999}}

\author{
 J.~Andruszkow$^{16}$, B.~Aune$^4$, V.~Ayvazyan$^{27}$, N.~Baboi$^{10}$, R.~Bakker$^2$,
 V.~Balakin$^3$, D.~Barni$^{14}$, A.~Bazhan$^{3}$, M.~Bernard$^{21}$, A.~Bosotti$^{14}$
% ...
}

\address{
1 Advanced Photon Source, Argonne National Laboratory, 9700 S.~Cass Avenue, 
Argonne, IL 60439, USA,\newline
2 BESSY, Albert-Einstein-Strasse 15, 12489 Berlin, Germany\newline
3 Branch of  the Inst.~of Nuclear Physics, 142284 Protvino, Moscow Region, Russia\newline
4 CEA Saclay, 91191 Gif s/Yvette, France\newline
5 Darmstadt University of Technology, FB18 - Fachgebiet TEMF, Schlossgartenstr.~8, 
64289 Darmstadt, Germany\newline
6 Deutsches Elektronen-Synchrotron DESY, Notkestrasse 85, 22603 Hamburg, Germany\newline
7 Deutsches Elektronen-Synchrotron DESY, Platanenallee 6, 15738 Zeuthen, Germany\newline
8 Faculty of Physics and Nuclear Techniques, University of Mining and Metallurgy, 
al.~Mickiewicza 30, PL-30-059 Cracow, Poland\newline
9 Fermi National Accelerator Laboratory, MS 306, P.O.~Box 500, Batavia, IL 60510 USA\newline
10 Hamburg University, Inst.~f.~Experimentalphysik, Notkestrasse 85, 20603 Hamburg, Germany\newline
11 Inst.~High Energy Physics IHEP, FEL Lab.~P.O.~Box 2732 Beijing 100080, P.R.~China\newline
12 Inst.~High Energy Physics, 142284 Protvino, Moscow Region, Russia\newline
13 INFN-LNF, via E.~Fermi 40, 00044 Frascati, Italy\newline
14 INFN Milano - LASA, via Fratelli Cervi 201, 20090 Segrate (MI), Italy\newline
15 INFN-Roma2, via della Ricerca Scientifica 1, 00100 Roma, Italy\newline
16 Institute of Nuclear Physics, Ul.~Kawiory 26 a, 30-55 Krakow, Poland\newline
17 Institute for Nuclear Research of RAS, 117312 Moscow, 60th October Anniversary prospect 7A, Russia\newline
18 Institute of Physics, Polish Academy of Sciences, al.~Lotnikow, 32/46, 02-668 Warsaw, Poland\newline
19 Institut de Physique Nucl\'{e}aire (CNRS-IN2P3), 91406 Orsay Cedex, France\newline
20  Joint Institute for Nuclear Research, 141980 Dubna, Moscow Region, Russia\newline
21 Laboratoire de l'Acc\'
{e}l\'{e}rateur Lin\'{e}aire, IN2P3-CNRS, Universit\'{e} de Paris-Sud, B.P.~34, F-91898 Orsay, France\newline
22 Max-Born-Institute, Max-Born-Str.~2a, 12489 Berlin, Germany\newline
23 Newman Lab, Cornell University, Ithaca, NY 14850, USA\newline
24 RWTH Aachen-Physikzentrum, Phys.~Inst.~IIIa, Sommerfeldstr.~26-28, 52056 Aachen, Germany\newline
25 UCLA Dept.~of Physics and Astronomy, 405 Hilgard Ave., Los Angeles, CA 90095, USA\newline
26 University of Rochester, Dept.~of Physics and Astronomy, 206 Bausch \& Lomb, 
Rochester NY 14627, USA\newline
27 Yerevan Physics Institute, 2 Alikhanyan Brothers str., 375036 Yerevan, Armenia\newline
\newline
\footnote[2]{}present address: CERN, CH 1211 Geneva 23, Switzerland\newline
\footnote[3]{}present address: Procter\&Gamble, 53881 Euskirchen, Germany\newline
\footnote[4]{}present address: Stanford Linear Accelerator Center, SLAC MS 07, 2575 Sand Hill Road, Menlo Park, 
CA 94025 USA\newline
\footnote[5]{}present address: Senderbetriebstechnik Westdeutscher Rundfunk, 50600 K\"oln, Germany \newline
\footnote[6]{}present address: UCLA Department of Physics \& Astronomy, Los Angeles, CA 90024, USA 
\newline 
\footnote[7]{}e-mail: joerg.rossbach@desy.de}

\setcounter{page}{0}

\maketitle

\begin{abstract} 
\newline
\noindent{\large \it Abstract} \newline
We present first observation of Self-Amplified Spontaneous Emission 
(SASE) in a free-electron laser (FEL) in the Vacuum Ultraviolet regime  
at 109~nm wavelength (11~eV). The observed free-electron laser gain 
(approx. 3000) and the radiation characteristics, such as dependency on 
bunch charge, angular distribution, spectral width and intensity 
fluctuations all corroborate the present models for SASE FELs.

\end{abstract}

\setcounter{page}{1}
\vspace*{-0.2cm}
\section{Introduction}
\vspace*{-0.5cm}
X-ray lasers are expected to open up new and exciting areas of basic 
and applied research in biology, chemistry and physics. Due to recent 
progress in accelerator technology the attainment of the long 
sought-after goal of wide-range tunable laser radiation in the 
Vacuum-Ultraviolet and X-ray spectral regions is coming close to 
realization with the construction of free-electron lasers (FEL) 
\cite{Madey} based on the principle of Self-Amplified Spontaneous 
Emission (SASE) \cite{Kondratenko,Bonifacio1}.  In a SASE FEL lasing 
occurs in a single pass of a relativistic, high-quality electron bunch 
through a long undulator magnet structure.

The radiation wavelength $\lambda_{\mathrm{ph}}$ of the first harmonic 
of FEL radiation is related to the period length $\lambda_{\mathrm{u}}$ 
of a planar undulator by

\begin{equation}
\lambda_{\mathrm{ph}}=
\frac{\lambda_{\mathrm{u}}}{2\gamma^2}\left(1+\frac{K^2}{2}\right)
\end{equation}

\noindent where $\gamma=E/(m_{\mathrm{e}}c^2)$ is the relativistic 
factor of the electrons, $K=eB_{\mathrm{u}}\lambda_{\mathrm{u}} /(2\pi 
m_{\mathrm{e}}c^2)$ the undulator parameter and $B_{\mathrm{u}}$ the 
peak magnetic field in the undulator. Equation (1) exhibits two main 
advantages of the free-electron laser: the free tunability of the 
wavelength by changing the electron energy and the possibility to 
achieve very short radiation wavelengths.

For most FELs presently in operation \cite{Colson}, the 
electron beam quality and the undulator length result in a gain of only 
a few percent per undulator passage, so that an optical cavity 
resonator and a synchronized multi-bunch electron beam have to be used. 
At very short wavelengths, normal-incidence mirrors of high 
reflectivity are unavailable. Therefore the generation of an electron 
beam of extremely high quality in terms of emittance, peak current and 
energy spread, and a high precision undulator of sufficient length are 
essential. Provided the spontaneous radiation from the first part of 
the undulator overlaps the electron beam, the electromagnetic radiation 
interacts with the electron bunch leading to a density modulation 
(micro-bunching) which enhances the power and coherence of radiation. 
In this ``high gain mode'' \cite{Kim,Krinsky,Saldin1}, the radiation 
power $P(z)$ grows exponentially with the distance $z$ along the 
undulator 

\begin{equation}
P(z)=AP_{\mathrm{in}}\exp(2z/L_{\mathrm{g}})
\end{equation}

\noindent where $L_{\mathrm{g}}$ is the field gain length, 
$P_{\mathrm{in}}$ the effective input power (see below), and $A$ the 
input coupling factor \cite{Krinsky,Saldin1}. $A$ is equal to 1/9 in 
one-dimensional FEL theory with an ideal electron beam. Typical 
parameters for a SASE FEL operating in the VUV wavelength range are: 
$P_{\mathrm{in}}$ of about a few Watts and power gain at saturation, $G 
= P_{\mathrm{sat}}/P_{\mathrm{in}}$, of about $10^8$.

Since the desired wavelength is very short, there is no laser tunable 
over a wide range to provide the input power $P_{\mathrm{in}}$.  
Instead, the spontaneous undulator radiation from the first part of the 
undulator is used as an input signal to the downstream part. FELs based 
on this Self-Amplified-Spontaneous-Emission (SASE) principle are 
presently considered the most attractive candidates for delivering 
extremely brilliant, coherent light with wavelength in the 
{\AA}ngstr\"om regime \cite{Winick,Brinkmann1,Brinkmann2,Nuhn}.  
Compared to state-of-the-art synchrotron radiation sources, one expects 
full transverse coherence, up to 4-6 orders of magnitude larger average 
brilliance, and up to 8-10 orders of magnitude larger peak brilliance 
at pulse lengths of about 200~fs FWHM.  Recently there have been 
important advances in demonstrating a high-gain SASE FEL at 12~$\mu$m 
wavelength \cite{Hogan} and at 530~nm wavelength \cite{Milton}.  

\vspace*{-0.2cm}
\section{Experimental set-up} 
\vspace*{-0.5cm}
The experimental results presented in this paper have been achieved at 
the TESLA Test Facility (TTF) Free-Electron Laser \cite{Brefeld} at the 
Deutsches Elektronen-Synchrotron DESY. The TESLA (TeV-Energy 
Superconducting Linear Accelerator) collaboration consists of 39 
institutes from 9 countries and aims at the construction of a 500~GeV 
(center-of-mass) e$^+$/e$^-$ linear collider with an integrated X-ray 
laser facility \cite{Brinkmann2}.  Major hardware contributions to TTF 
have come from Germany, France, Italy, and the USA.  The goal of the 
TTF FEL is to demonstrate SASE FEL emission in the VUV and, in a second 
phase, to build a soft X-ray user facility \cite{Aberg,Rossbach}. 

\begin{table}
\caption{
Main parameters of  the TESLA Test Facility for FEL experiments (TTF 
FEL, phase 1)}
\begin{tabular}{|lc|}
\bf{Parameter} &  \bf{Measured value} \\
\tableline
beam energy                            &  $233 \pm 5$ MeV\\
rms energy spread                      &  $0.3 \pm 0.2$ MeV \\
rms transverse beam size               &  $100 \pm 30 \ \mu $m \\
normalized emittance, $\varepsilon_{\mathrm{n}}$ 
&  $6 \pm 3 \ \pi$ mrad mm \\
electron bunch charge &  1 nC\\
peak electron current &  $400 \pm 200$ A \\
bunch spacing &  1 $\mu $s\\
repetition rate &  1 Hz\\
\tableline
undulator period, $\lambda_{\mathrm{u}}$  &  27.3 mm \\
undulator peak field &  0.46 T\\
effective undulator length &  13.5 m\\
\tableline
radiation wavelength, $\lambda_{\mathrm{ph}}$ 
&  109 nm \\
FEL gain &  $ 3 \times 10^3$ \\
FEL radiation pulse length &  0.3-1 ps \\
\end{tabular}
\label{table}
\end{table}

\begin{figure}
\epsfig{file=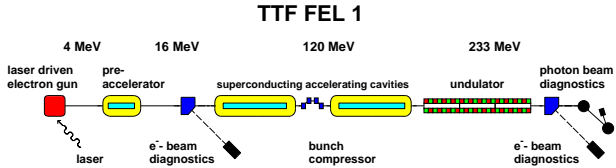,angle=270,width=0.45\textwidth}
\vspace*{3mm}
\caption{              
Schematic layout of phase 1 of the SASE FEL at the TESLA Test Facility 
at DESY, Hamburg. The linac contains two 12.2 m long cryogenic modules 
each equipped with eight 9-cell superconducting accelerating cavities 
[20]. The total length is 100 m.
} 
\label{layout}
\end{figure}

The layout is shown in Fig.~\ref{layout}. The main parameters for FEL 
operation are compiled in Table~\ref{table}.
The injector is based on a laser-driven 1$\frac{1}{2}$-cell rf gun 
electron source operating at 1.3~GHz \cite{Carneiro}.  It uses a 
Cs$_2$Te cathode \cite{Michelato} and can generate bunch charges more 
than 10~nC at 1~MHz repetition rate.  A loading system allows mounting 
and changing of cathodes while maintaining ultra-high vacuum conditions 
\cite{Michelato}.  The cathode is illuminated by a train of UV laser 
pulses generated in a mode-locked solid-state laser system \cite{Will} 
synchronized with the rf. An energy of up to 50~$\mu$J with a 
pulse-to-pulse variation of 2\% (rms) is achieved. The 
UV pulse length measured with a streak camera is $\sigma_{\mathrm{t}} = 
7.1 \pm 0.6$~ps . The rf gun is operated with a peak electric field of 
37~MV/m on the photocathode. The rf pulse length was limited to 
100~$\mu$s and the repetition  rate to 1~Hz for machine protection 
reasons.  The gun section is followed by a 9-cell superconducting 
cavity, boosting the energy to 16 MeV. The superconducting accelerator 
structure has been described elsewhere \cite{Weise}.

The undulator is a fixed 12~mm gap permanent magnet device using a 
combined function magnet design \cite{Nikitina} with a period length of 
$\lambda_{\mathrm{u}}$ = 27.3~mm and a peak field of $B_{\mathrm{u}}$ = 
0.46~T, resulting in an undulator parameter of $K=1.17$. Integrated 
quadrupole structures produce a gradient of 12~T/m superimposed on the 
periodic undulator field in order to focus the electron beam along the 
undulator. The undulator system is subdivided into three segments, each 
4.5~m long and containing 10 quadrupole sections to build up 5 full 
focusing-defocusing (FODO) cells. The FODO lattice periodicity runs 
smoothly from segment to segment. There is a spacing of 0.3~m between 
adjacent segments for diagnostics.  The total length of the system is 
14.1~m. The vacuum chamber incorporates 10 beam position monitors and 
10 orbit correction magnets per segment, one for each quadrupole 
\cite{Hahn1,Hahn2}. 

For optimum overlap between the electron light beams, high precision on 
the magnetic fields and mechanical alignment are required. The 
undulator field was adjusted such that the expected rms deviations of 
the electron orbit should be smaller than 10~$\mu$m at 300~MeV 
\cite{Pflueger1}. The beam orbit straightness in the undulator is 
determined by the alignment precision of the superimposed 
permanent-magnet quadrupole fields which is better than 50~$\mu$m in 
both vertical and horizontal direction. The relative alignment of the 
three segments is accomplished with a laser interferometer to better 
than 30~$\mu$m \cite{Pflueger2}.  

Different techniques have been used to measure the emittance of the 
electron beam \cite{Edwards}: Magnet optics scanning (``quadrupole 
scans''), tomographic reconstruction of the phase space including space 
charge effects, and the slit system method. All methods use optical 
transition radiation emitted from aluminum foils to measure the bunch 
profiles and yield values for the normalized emittance of ($4 \pm 
1$)~$\pi$ mrad mm for a bunch charge of 1~nC at the exit of the 
injector.  The emittance in the undulator, as determined from 
quadrupole scans and from a system of wire scanners was typically 
between 6 and 10~$\pi$~mrad~mm (in both horizontal and vertical phase 
space).  It should be noted that the measurement techniques applied 
determine the emittance integrated over the entire bunch length. 
However, for FEL physics, the emittance of bunch slices much shorter 
than the bunch length is the relevant parameter. It is likely that, due 
to spurious dispersion and wakefields, the bunch axis is tilted about a 
 transverse axis such that the projected emittance is larger than the 
emittance of any slice.  Based on these considerations we estimate the 
normalized slice emittance in the undulator at ($6 \pm 
3$)~$\pi$~mrad~mm. 

A bunch compressor is inserted between the two accelerating modules, in 
order to increase the peak current of the bunch up to 500~A, 
corresponding to 0.25~mm bunch length (rms) for a 1~nC bunch with 
Gaussian density profile.  Experimentally, it is routinely verified 
that a large fraction of the bunch charge is compressed to a length 
below 0.4~mm (rms) \cite{Geitz}. There are indications that the core is 
compressed even further.  We estimate the peak current for the FEL 
experiment at ($400 \pm 200$)~A. Coherent synchrotron radiation in the 
magnetic bunch compressor may affect the emittance and the energy 
spread at such short bunch lengths \cite{Dohlus}. 

\section{FEL measurements} 

\begin{figure}
\epsfig{file=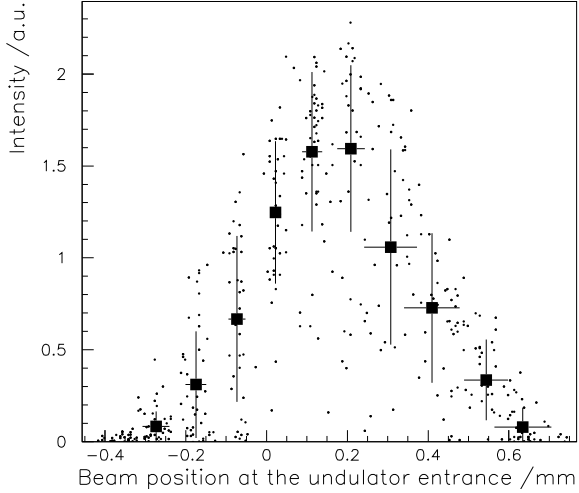,angle=0,width=0.43\textwidth}
\vspace*{2mm}
\caption{
Sensitivity of radiation power to horizontal electron beam position at 
the undulator entrance. The dots represent mean values of the radiation 
intensity for each beam position. The horizontal error bars denote the 
rms beam position instability while the vertical error bars indicate 
the standard deviation of intensity fluctuations, which are due to the 
statistical charcter of the SASE process, see. Eq. (3)} 
\label{position}
\end{figure}

A strong evidence for the FEL process is a large increase in the 
on-axis radiation intensity if the electron beam is injected such that 
it overlaps with the radiation during the entire passage through the 
undulator. Fig. \ref{position} shows the intensity passing a 0.5~mm 
iris, located on axis 12~m downstream of the undulator, as a function 
of the horizontal beam position at the undulator entrance. The observed 
intensity inside a window of $\pm 200 \ \mu$m around the optimum beam 
position is a factor of more than 100 higher than the intensity of 
spontaneous radiation. A PtSi photodiode was used integrating over all 
wavelengths. Note that the vacuum chamber diameter in the undulator 
(9.5~mm) is much larger than the beam diameter (300~$\mu$m).

\begin{figure}
\epsfig{file=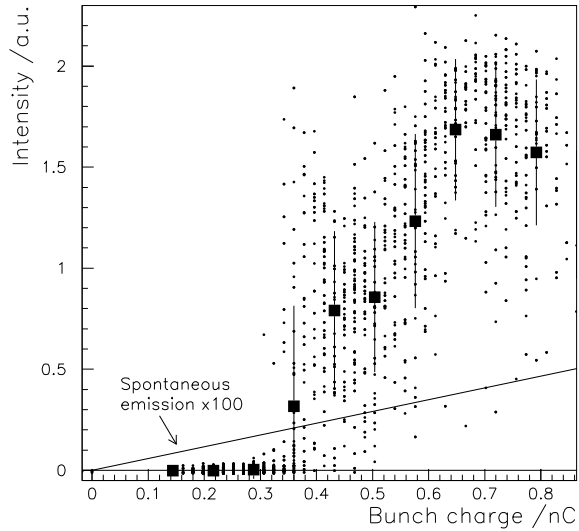,angle=0,width=0.43\textwidth}
\vspace*{2mm}
\caption{
SASE intensity versus bunch charge. The straight line is the 
spontaneous intensity multiplied by a factor of 100.  To guide the eye, 
mean values of the radiation intensity are shown for some bunch charges 
(dots). For vertical error bars, see Fig. 2}
\label{charge}
\end{figure}

SASE gain is expected to depend on the bunch charge in an extremely 
nonlinear way. Fig.~\ref{charge} shows the measured intensity on axis 
as a function of bunch charge $Q$, while the beam orbit is kept 
constant for optimum gain. The solid line indicates the intensity of 
the spontaneous undulator radiation multiplied by a factor of 100. The 
strongly nonlinear increase of the intensity as a function of bunch 
charge is a definite proof of FEL action. The gain does not further 
increase if the bunch charge exceeds some 0.6~nC. This needs further 
study, but it is known that the beam emittance becomes larger for 
increasing $Q$ thus reducing the FEL gain.

\begin{figure}
\epsfig{file=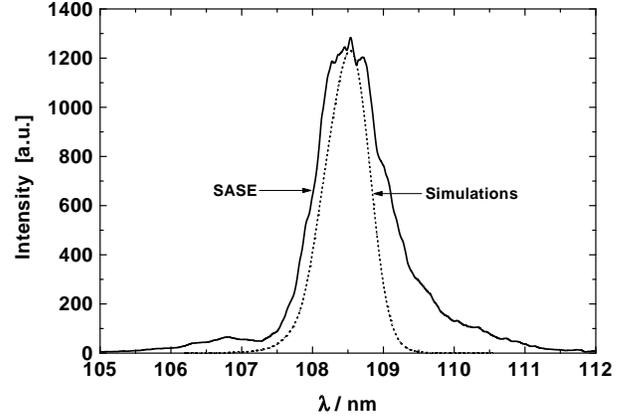,angle=0,width=0.45\textwidth}
\vspace*{3mm}
\caption{
Wavelength spectrum of the central radiation cone (collimation angle 
$\pm$0.2 mrad), taken at maximum FEL gain. The dotted line is the 
result of numerical simulation. The bunch charge is 1 nC.}
\label{spectrum}
\end{figure}

\begin{figure}
\epsfig{file=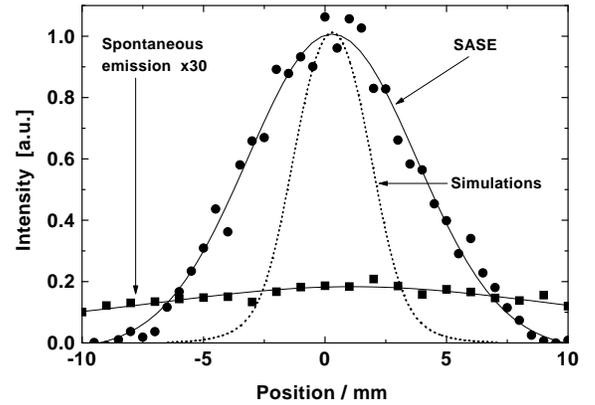,angle=0,width=0.45\textwidth}
\vspace*{3mm}
\caption{
Horizontal intensity profile of SASE FEL and spontaneous undulator 
radiation ($x30$), measured with a photodiode behind a 0.5 mm aperture 
in a distance of 12 m from the end of the undulator. The dotted line is 
the result of numerical simulation.
} 
\label{horprofile}
\end{figure}

The wavelength spectrum of the radiation (taken on axis at maximum FEL 
gain) is presented in Fig.~\ref{spectrum}.  The central wavelength of 
108.5~nm is consistent with the measured beam energy of ($233 \pm 
5$)~MeV and the known undulator parameter $K=1.17$, see Eq. (1). The 
intensity gain determined with the CCD camera of the spectrometer is in 
agreement with the photodiode result.

A characteristic feature of SASE FELs is the concentration of radiation 
power into a cone much narrower than that of wavelength integrated 
undulator radiation, whose opening angle is in the order of 
$1/\gamma $. Measurements done by moving the 0.5~mm iris horizontally 
together with the photodiode confirm this expectation, see 
Fig.~\ref{horprofile}. The spontaneous intensity is amplified by a 
factor of 30 to be visible on this scale.  

In order to study which section of the undulator contributes most to 
the FEL gain, we applied closed orbit beam bumps to different sections 
of the undulator, thus disturbing the gain process at various 
locations along the undulator. It was seen that practically the entire 
undulator contributes, but with some variation in local gain.  Some 
improvement in the over-all gain should be possible by optimizing the 
settings of the 30 orbit correction coils which have not been touched 
so far.

To determine the absolute FEL intensity, the photodiode is firstly 
calibrated using the known intensity of undulator radiation. For the 
iris diameter of 3~mm used for spontaneous emission and averaging the 
solid angle $\Omega$ over the source positions inside the undulator, we 
get $\Omega = 2\times 10^{-8}$. About 70~pJ is accepted within the iris 
at the first harmonic of the spontaneous undulator radiation per 1~nC 
bunch charge. The SASE FEL signal was measured using an iris diameter 
of 0.5~mm, the signal was larger by a factor of about 5. Hence, the 
energy flux is 2~nJ/mm$^2$ at the location of the detector and the 
on-axis flux per unit solid angle is about 0.3~J/sr (assuming a source 
position at the end of the undulator). This value was used as input for 
the numerical simulation of the SASE FEL performed with the code FAST 
\cite{Saldin2}. The longitudinal profile of the bunch current was 
assumed to be Gaussian with an rms length of 0.25~mm. The transverse 
distribution of the beam current density was also taken to be Gaussian.  
Calculations have been performed for a Gaussian energy spread of 
0.1\%. According to numerical simulation one of the most critical 
parameters for FEL operation is the normalized emittance 
$\varepsilon_n$ that was varied in the simulations between 5 and 
10~$\pi$~mrad~mm. A first conclusion from our calculations is that the 
TTF FEL operates in the high-gain linear regime where the power grows 
exponentially along the undulator. The contribution of the fundamental 
transverse mode TEM$_{00}$ to the total power seems to dominate, so 
that Eq. (2) applies. 

Our calculations show that the length at which a level of energy flux 
of 0.3~J/sr is obtained strongly depends on emittance, but the number 
of gain lengths is roughly the same in all cases and is about 5.
Figs.~\ref{spectrum} and \ref{horprofile} include typical theoretical 
spectral and angular distributions as calculated by our numerical 
simulation. In both cases experimental curves are wider than the 
simulation results. A possible source of the widening is energy and 
orbit jitter, since the experimental curves are results of averaging 
over many bunches. 

The FEL gain is defined as the ratio of output to input 
power$P_{\mathrm{out}}/P_{\mathrm{in}}$, see Eq. (2). It is a 
characteristic of the FEL amplifier and should depend only on the 
parameters of the electron beam and the undulator but not on the type 
of input signal.  For a FEL amplifier seeded by an external laser the 
input power is well defined.  For an FEL amplifier starting from noise 
(i.e. a SASE FEL) the effective power of shot noise can be defined as 
the power of optimally focused seeding radiation yielding the same 
output power. The gain $G$ is then simply $G = A 
\exp(2z/L_{\mathrm{g}})$, see Eq.  (2). With an input coupling factor A 
$\approx$ 0.1, the FEL gain can be estimated at $G \approx 3 \cdot 
10^3$.  The uncertainty is estimated at a factor of 3 (i.e. $10^3 < G 
< 10^4$) and is mainly due to the imprecise knowledge of the 
longitudinal beam profile.

It is essential to realize that the fluctuations seen in 
Figs.~\ref{position} and \ref{charge} are not primarily due to unstable 
operation of the accelerator but are inherent to the SASE process. Shot 
noise in the electron beam causes fluctuations of the beam density, 
which are random in time and space \cite{Bonifacio2}. As a result, the 
radiation produced by such a beam has random amplitudes and phases in 
time and space and can be described in terms of statistical optics. In 
the linear regime of a SASE FEL, the radiation pulse energy measured in 
a narrow central cone (opening angle $\pm 20 \ \mu$rad in our case) at 
maximum gain is expected to fluctuate according to a gamma distribution 
$p(E)$ \cite{Saldin3},

\begin{equation}
p(E)=\frac{M^M}{\Gamma(M)}\left(\frac{E}{\langle E \rangle}\right)^{M-1}
\frac{1}{\langle E \rangle }
\exp\left(-M\frac{E}{\langle E \rangle}\right)
\end{equation}

\noindent where $\langle E \rangle$ is the mean energy, $\Gamma(M)$ is 
the gamma function with argument $M$, and $M^{-1} = \langle(E-\langle E 
\rangle)^2\rangle /\langle E\rangle ^2$ is the normalized variance of 
$E$. The parameter $M$ corresponds to the number of 
longitudinal optical modes.  Note that the same kind of statistics 
applies for completely chaotic polarized light, in particular for  
spontaneous undulator radiation. 

\begin{figure}
\epsfig{file=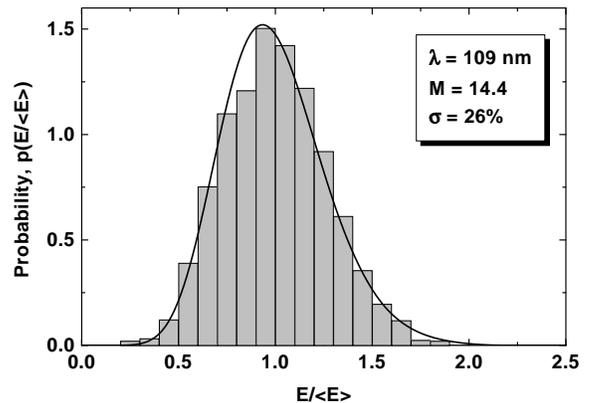,angle=0,width=0.45\textwidth}
\vspace*{3mm}
\caption{
Probability distribution of SASE intensity. The rms fluctuation yields 
a number of longitudinal modes $M = 14$. The solid curve is the gamma 
distribution for $M = 14.4$. The bunch charge is 1~nC.
} 
\label{gamma}
\end{figure}

For these statistical measurements the signals from 3000 radiation 
pulses have been recorded, with the small iris (0.5~mm diameter) in 
front of the photo diode to guarantee that transversely coherent 
radiation pulses are selected. As one can see from Fig.~\ref{gamma}, 
the distribution of the energy in the radiation pulses is quite close 
to the gamma distribution. The relative rms fluctuations are about 26\% 
corresponding to $M = 14.4$. One should take into account that these 
fluctuations arise not only from the shot noise in the electron beam, 
but the pulse-to-pulse variations of the beam parameters can also 
contribute to the fluctuations. Thus, the value $M \approx 14$ can be 
considered as a lower limit for the number of longitudinal modes in the 
radiation pulse. Using the width of radiation spectrum we calculate the 
coherence time \cite{Saldin3} and find that the part of the electron 
bunch contributing to the SASE process is at least 100~$\mu$m long. 
>From the quality of the agreement with the gamma distribution we can 
also conclude that the statistical properties of the radiation are 
described with Gaussian statistics.  In particular, this means that 
there are no FEL saturation effects.

\acknowledgments

Fundamental work on high accelerating gradients in superconducting 
cavities at Wuppertal University was essential for the successful 
construction of the first TTF accelerating modules. We are grateful for 
the invaluable support by the technical staff of the participating 
groups. Support by the Moscow Physical Engineering Institute and by 
the Institute for Nuclear Studies, Swierk, Poland, is gratefully 
acknowledged.


\begin{references}

\bibitem{Madey}
J.M.J.~Madey, J. Appl. Phys. {\bf 42,} 1906 (1971).

\bibitem{Kondratenko}
A.M.~Kondratenko, E.L.~Saldin,  Part. Accelerators {\bf 10,} 207 (1980)

\bibitem{Bonifacio1}
R.~Bonifacio, C.~Pellegrini, L.M.~Narducci, Opt. Commun. {\bf 50,} 373 (1984)

\bibitem{Colson}
W.B.~Colson, Nucl. Instr. and Meth. {\bf A429,} 37-40 (1999).

\bibitem{Kim}
K.J.~Kim, Phys. Rev. Lett. {\bf 57,} 1871 (1986)

\bibitem{Krinsky}
S.~Krinsky, L.H.~Yu, Phys. Rev. {\bf A35,} 3406 (1987)

\bibitem{Saldin1}
E.L.~Saldin, E.A.~Schneidmiller, M.V.~Yurkov, ``The Physics of Free Electron Lasers'', Springer (1999) and references therein

\bibitem{Winick}
H.~Winick, et al., Proc. PAC Washington and SLAC-PUB-6185, (1993)

\bibitem{Brinkmann1}
R.~Brinkmann, et al., Nucl. Instr. and Meth. {\bf A 393,} 86-92 (1997)

\bibitem{Brinkmann2}
R.~Brinkmann, G.~Materlik, J.~Rossbach, A.~Wagner (eds.), DESY 1997-048 and ECFA 1997-182 (1997)

\bibitem{Nuhn}
H.-D.~Nuhn, J.~Rossbach, Synchrotron Radiation News {\bf 13}, No. 1, 18 - 32 (2000)

\bibitem{Hogan}
M.~Hogan, et al., Phys. Rev. Lett. {\bf 81,} 4867 (1998)

\bibitem{Milton}
S.~Milton, et al., ``Observation of Self-Amplified Spontaneous Emission and Exponential Growth at 530 nm'', 
submitted to Phys. Rev. Lett. (2000)

\bibitem{Brefeld}
W.~Brefeld, et al., Nucl. Instr. and Meth. {\bf A393,}  119-124 (1997)

\bibitem{Aberg}
T.~{\AA}berg, et al., A VUV FEL at the TESLA Test Facility at DESY, Conceptual Design Report,  DESY Print 
TESLA-FEL 95-03 (1995)

\bibitem{Rossbach}
J.~Rossbach, Nucl. Instr. and Meth. {\bf A 375,} 269 (1996)

\bibitem{Carneiro}
J.-P.~Carneiro, et al., Proc. 1999 Part. Acc. Conf., New York, 2027-2029 (1999)

\bibitem{Michelato}
P.~Michelato, et al., 
Nucl. Instr. and Meth. {\bf A445,} 422 (2000)

\bibitem{Will}
I.~Will, S.~Schreiber, A.~Liero, W.~Sandner, 
Nucl. Instr. and Meth. {\bf A445,} 427 (2000)

\bibitem{Weise}
H.~Weise, Proc. 1998 Linac Conf. Chicago, 674-678 (1998)

\bibitem{Nikitina}
Y.M.~Nikitina, J. Pfl\"uger, Nucl. Instr. and Meth. {\bf A375,} 325 
(1996)

\bibitem{Hahn1}
U.~Hahn, J.~Pfl\"uger, G.~Schmidt, Nucl. Instr. and Meth. {\bf A429,} 
276 (1999)

\bibitem{Hahn2}
U.~Hahn, et al., 
Nucl. Instr. and Meth. {\bf A445,} 442 (2000)

\bibitem{Pflueger1}
J.~Pfl\"uger, 
Nucl. Instr. and Meth. {\bf A445,} 366 (2000)

\bibitem{Pflueger2}
J.~Pfl\"uger, H.~Lu, T.~Teichmann, Nucl. Instr. and Meth. {\bf A429} 
386 (1999)

\bibitem{Edwards}
H.~Edwards, et al., Proc. 1999 FEL Conf., Hamburg

\bibitem{Geitz}
M.~Geitz, G.~Schmidt, P.~Schm\"user, G.V.~Walter, 
Nucl. Instr. and Meth. {\bf A445,} 343 (2000)

\bibitem{Dohlus}
M.~Dohlus, A.~Kabel, T.~Limberg, Proc. 1999 Part. Acc. Conf., New York, 1650-1652 (1999)

\bibitem{Saldin2}
E.L.~Saldin, E.A.~Schneidmiller, M.V.~Yurkov, Nucl. Instr. and Meth. {\bf A 429,} 233 (1999) 

\bibitem{Bonifacio2}
R.~Bonifacio, et al., Phys. Rev. Lett. {\bf 73,} 70 (1994)

\bibitem{Saldin3}
E.L.~Saldin, E.A.~Schneidmiller, M.V.~Yurkov, Opt. Commun. {\bf 148} 383 (1998)

\end{references}
\end{document}